\documentstyle[aps,preprint,epsfig,eqsecnum]{revtex}
\begin{document}
\draft
\title{Modeling of Nucleon--Nucleon Potentials,\\
Quantum Inversion versus Meson Exchange Pictures\footnote{Contribution
to the International Conference on Inverse and Algebraic Quantum
Scattering Theory, Lake Balaton '96.}}
\author{L. J\"ade, M. Sander and H.\,V. von Geramb}
\address{Theoretische Kernphysik, Universit\"at Hamburg\\
Luruper Chaussee 149, D-22761 Hamburg}
\date{\today}
\maketitle
\begin{abstract}
The notion of interacting elementary particles for low and medium energy
nuclear physics is associated with definitions of potential operators
which, inserted into a Lippmann--Schwinger equation, yield the scattering
phase shifts and observables. In principle, this potential carries the
rich substructure consisting of quarks and gluons and thus may be
deduced from some microscopic model. In this spirit we propose a boson
exchange potential from a nonlinear quantum field theory. Essentially,
the meson propagators and form factors of conventional models are
replaced by amplitudes derived from the dynamics of self--interacting
mesons in terms of solitary fields. Contrary to deduction, 
we position the inversion
approach. Using Gel'fand--Levitan and Marchenko inversion we compute
local, energy--independent potentials from experimental phase shifts for
various partial waves. Both potential models give excellent results for
on--shell NN scattering data. In the off--shell domain we study both
potential models in ($p$, $p\gamma$) Bremsstrahlung, elastic
nucleon--nucleus scattering and triton binding energy calculations. It
remains surprising that for all observables the inversion and
microscopic meson exchange potentials are equivalent in their
reproduction of data. Finally, we look for another realm of elementary
interactions where inversion and meson exchange models can be applied
with the hope to find more sensitivity to discern substructure dynamics.  
\end{abstract}
\section{Introduction}
In the last decade, the development of highly quantitative models for the
nucleon--nucleon interaction was one of the major tasks of theoretical
nuclear physicists. Based on Yukawa's pioneer work, 
potentials for NN forces mediated by 
the exchange of boson fields with masses below 1\,GeV were invented and
successfully applied to various problems in medium energy nuclear
physics \cite{pot}. Despite of their remarkable ability to account for
quantitative details of NN phenomenology, none of these models contains
any reference to Quantum Chromodynamics QCD, which is believed to be the
underlying microscopic theory of the strong interaction. There are a
number of models which explicitly refer to QCD \cite{chi}, 
but so far all of them
fail to describe the nucleon--nucleon interaction comparable well as the
phenomenological boson exchange potentials. 
Thus the major shortcomings of today's nucleon--nucleon potential models
are the empirical character of the boson--exchange potentials which
arises from phenomenological usage of form factors without 
stringent connections to QCD and on the other side the failure of QCD
inspired models to provide a quantitative description of
nucleon--nucleon scattering data. 
  
The goal of the Solitary Boson Exchange Potential, which will be
described in detail in section \ref{osbep} is to interpolate
between these extreme positions \cite{jae96}. 
Characteristic features of QCD inspired
potential models, namely their nonlinear character, are taken into
account using a nonlinear expansion of the Klein--Gordon equation as
equation of motion for the boson fields. On the other hand, the
solutions of this equation, called {\it solitary boson fields}, are used
in a boson exchange potential which is in great analogy to the Bonn--B
potential \cite{Mach89} to obtain a quantitative description of NN data
comparable well as the phenomenological boson exchange potentials. 
It will turn out that the nonlinear character of the boson
fields allows to substitute the phenomenological form factors of the
Bonn potential. Simultaneously, an empirical scaling law is found which
relates all meson parameters, as expected in a model based on QCD. 
  
Contrary to the microscopic models, potentials obtained from quantum 
inversion were developed for various 
hadron--hadron interactions \cite{inv}. Using experimental phase shifts
as input in the Gel'fand--Levitan and Marchenko
inversion algorithm for the Sturm--Liouville equation yields 
local potential operators in coordinate space  
which are model--independent and
reproduce the experimental data by construction. 
  
As far as elastic NN data are concerned, inversion potentials evidently 
provide a precise description of the NN interaction. It is a nontrivial
question, however, whether this accuracy remains in the application of
inversion potentials to more complex problems. Since the scattering
phase shifts, from which inversion potentials are obtained, 
only contain the {\it on--shell} information of the scattering
amplitude, i.\,e.\ the absolute value of the incoming and outgoing
nucleon momentum remains unchanged, it is questionable if such a
potential can account for the description of reactions where the {\it
off--shell} part of the $t$--matrix contributes. A sensitivity on
details of the off--shell amplitude would provide the desired
possibility to test possible effects of the substructure of potential
models. In particular, we seek a signature in the data to confirm the
assumptions which led to the Solitary--Boson--Exchange--Potential.   
To study this
interesting point we apply boson exchange as well as inversion
potentials to calculate the differential
cross section for ($p$, $p\gamma$) Bremsstrahlung and elastic
nucleon--nucleus scattering as well as the triton binding energy. The 
astonishing outcome implies that inversion and boson
exchange potentials yield equivalent results. Even more surprising, an
improvement of the description of the on--shell data enhances the
accuracy describing the off--shell data.   
  
Before inversion and boson exchange potentials will be compared, we
give a short reminder of the algorithms which 
are used to calculate inversion
potentials from experimental phase shifts and show the typical structure
of boson exchange potentials outlining the basic ideas of our 
Solitary--Boson--Exchange--Potential model.  
\section{Nucleon--Nucleon Potentials from Inversion}
\label{inv}
Contrary to the direct path to obtain a potential for NN scattering from
some microscopic model, the algorithm of quantum inversion can be applied 
using experimental phase shifts as input
\cite{inv}. Nowadays, the inversion techniques for nucleon--nucleon
quantum scattering have evolved up to almost perfection for scattering
data below pion production threshold. Numerically, input phase shifts
can be reproduced for single and coupled channels 
with a precision of $1/100$ of a degree, which is much
lower than the experimental uncertainty. This accuracy and the
possibility to test the inversion potential `online', i.\,e.\ inserting the
potential into the scattering equation to reproduce the input phase
shifts, makes them a reliable and easy--to--handle tool for highly
quantitative medium energy nuclear physics. Guided by this spirit, the
utmost aim of quantum inversion is to provide the most simple operator
to reproduce the scattering data. This paradigm, however, proscribes to
include any momentum dependence and thus any non--locality in the
potential since this would open a box full of ambiguities which can not
be associated with the goal of simplicity. 
   
As a basis, the radial Schr\"odinger equation is assumed to be the
relevant equation of motion for the two--particle system
\begin{equation}
\left\{ -{d^2 \over dr^2} + {\ell ( \ell+1) \over r^2 } + 
{2 \mu \over \hbar^2} V_\ell (r)
\right\} \psi_\ell (k,r) = k^2 \psi_\ell (k,r),
\label{rse}
\end{equation}
where $V_\ell (r)$ is a local, energy--independent operator in
coordinate space. Substituting
\begin{equation}
 q(r)  =  
{ \ell ( \ell +1) \over r^2} + { 2 \mu \over \hbar^2} V_\ell
(r)\qquad\mbox{and}\qquad \lambda=k^2,
\end{equation}
one obtains the well--known Sturm--Liouville equation
\begin{equation}
\left[ - {d^2 \over dx^2} + q(x) \right] y (x) = \lambda y(x).
\end{equation}
The scattering phase shifts enter as boundary conditions for the
physical solutions of (\ref{rse}) which read
\begin{equation} \lim_{r \rightarrow \infty} \psi_\ell (k,r) =
\exp ( i {\delta_\ell (k)}) \sin (kr - {\ell \pi \over 2} 
+ {\delta_\ell (k)})
\end{equation}
There are two equivalent inversion algorithms for the Sturm--Liouville
equation, the Marchenko and the Gel'fand--Levitan inversion, which will 
be outlined in the next sections for the case of uncoupled channels. 
\subsection{Marchenko Inversion}
In the Marchenko inversion the experimental information enters via the
$S$--matrix, which is related to the scattering phase shifts by the
simple relation
\begin{equation}
\label{sm}
S_{\ell}(k)=\exp(2i\delta_{\ell}(k)).
\end{equation}
Inserting a rational representation of the $S$--matrix \cite{inv} into
the integral equation for the input kernel
\begin{equation} \label{minpk}
F_\ell (r,t) = -\frac{1}{2\pi} \int_{-\infty}^{+\infty} h^+_\ell(kr)
                              \left[ S_\ell(k)-1
 \right] h^+_\ell(kt) dk,
\end{equation} 
where $h^+_\ell(x)$ are the Riccati--Hankel functions, 
the Marchenko equation 
\begin{equation} \label{mfeqn}
A_\ell (r,t)+F_\ell (r,t)+\int_{r}^{\infty}A_\ell(r,s)F_\ell(s,t)ds = 0,
\end{equation}
becomes an algebraic equation for the translation kernel
$A_{\ell}(r,t)$. The potential is obtained by taking the derivative 
\begin{equation}
\label{potm}
V_{\ell}(r)=-2\frac{d}{dr}A_{\ell}(r,r).
\end{equation}
\subsection{Gel'fand--Levitan Inversion}
Instead of the $S$--matrix, in the Gel'fand--Levitan inversion 
the Jost--matrix carries the experimental input. The latter is related
to the $S$--matrix by
\begin{equation}
\label{rhp}
S_{\ell}(k)=\frac{F_{\ell}(-k)}{F_{\ell}(k)},
\end{equation}
and leads to the input kernel
\begin{equation}
G_\ell (r,t) = \frac{2}{\pi} \int_{0}^{\infty} j_\ell (kr) \left[
             \frac{1}{|F_\ell (k)|^2}- 1
 \right] j_\ell (kt) dk,
\end{equation}
where $j_{\ell}(x)$ are the Riccati--Bessel functions. 
A rational representation of the spectral density \cite{inv} again
yields an algebraic form for the Gel'fand--Levitan equation
\begin{equation}
K_\ell (r,t)+G_\ell (r,t)+\int_{0}^{r}K_\ell (r,s)G_\ell (s,t)ds = 0,
\end{equation}
and the desired potential is obtained from
\begin{equation}
\label{vgl}
V_{\ell}(r)=2\frac{d}{dr}K_{\ell}(r,r).
\end{equation}
\subsection{Coupled Channel Inversion}
For coupled channels, i.\,e.\ transitions between states with different
angular momentum $\ell_i$, the Schr\"odinger equation (\ref{rse}) becomes a
matrix equation. 
\begin{equation}
\label{mrse}
\left\{-\frac{d^2}{dr^2}{\bf 1} + \mbox{\bf  V}(r)\right\}{\bf
\Psi}(r)=k^2{\bf \Psi}(r),
\end{equation}
where in the case of two coupled angular momentum the potential matrix
reads 
$$
\mbox{\bf  V}(r)=\left(\begin{array}{cc}
\displaystyle{\frac{\ell_1(\ell_1+1)}{r^2}+
\frac{2\mu}{\hbar^2}V_{\ell_1\ell_1}(r)} &
\displaystyle{\frac{2\mu}{\hbar^2}V_{\ell_1\ell_2}(r)} \\
 & \\
\displaystyle{\frac{2\mu}{\hbar^2}V_{\ell_2\ell_1}(r)} & 
\displaystyle{\frac{\ell_2(\ell_2+1)}{r^2}+
\frac{2\mu}{\hbar^2}V_{\ell_2\ell_2}(r)} 
\end{array}\right).
$$
The input and translation kernels of the previous sections 
and in particular the $S$--matrix now  generalize to matrices. Since
it is cumbersome for coupled channel 
situations to solve the Riemann--Hilbert
Problem (\ref{rhp}) numerically we will focus on the Marchenko inversion
which does not include any serious difficulty for the general case of
coupled channels. Defining the diagonal matrix which contains the
Riccati--Hankel functions by
$$
\mbox{\bf  H}(x) = \left(\begin{array}{cc} h_{\ell_1}^+(x) & 0 \\
 & \\ 0  & h_{\ell_2}^+(x) \end{array}\right),
$$
one gets as a generalization of (\ref{minpk}) for the input kernel
\begin{equation}
\label{minpkm}
\mbox{\bf F} (r,t) = -\frac{1}{2\pi} \int_{-\infty}^{+\infty} \mbox{\bf
H}(kr) \left[ \mbox{\bf S}(k)-{\bf 1}
 \right]\mbox{\bf H}(kt)  dk + \sum_{i=1}^{N_B}\mbox{\bf
H}(k_ir)\mbox{\bf N}(k_i)\mbox{\bf H}(k_it),
\end{equation} 
where the matrix {\bf N}($k_i$) contains the asymptotic normalizations
of the wave functions for the bound states at (imaginary) momentum
$k_i$. The Marchenko fundamental equation (\ref{mfeqn}) now reads
\begin{equation} \label{mfeqnm}
\mbox{\bf A} (r,t)+\mbox{\bf F} (r,t)+\int_{r}^{\infty}\mbox{\bf
A}(r,s)\mbox{\bf F}(s,t)ds = 0,
\end{equation}
and the potential matrix is obtained from
\begin{equation}
\mbox{\bf V}(r)=-2\frac{d}{dr}\mbox{\bf A} (r,r).
\end{equation}
\section{The One--Solitary--Boson--Exchange Potential}
\label{osbep}
In the following section, our Solitary--Meson--Exchange--Potential model
is used to demonstrate the major problems in the modeling of
nucleon--nucleon potentials using the boson exchange picture. As
mentioned in the introduction, today's models for NN interaction either
refer to QCD but can not describe scattering data or --- as is the case
for the conventional boson exchange potentials --- they 
contain empirical entities which must be fitted to experiment. 
  
The One--Solitary--Boson--Exchange--Potential OSBEP which was recently
developed by the Hamburg group \cite{jae96} 
tries to interpolate between these
extreme positions. Surprisingly, it turns out that the inclusion of
typical features from QCD inspired models can account for the empirical
parts of usual boson exchange potentials preserving the high accuracy in 
the description of data. Therefore, we take the OSBEP as an
illustrative example of how a microscopic model is used to obtain a
boson exchange potential for nucleon--nucleon interactions.
\subsection{Solitary Mesons}
Motivated by the nonlinear character of QCD, we 
assume a nonlinear self--interaction for {\it all} mesons which enter
the boson exchange potential. Doing so, we use the model of solitary
mesons 
developed by Burt \cite{Burt81}. Here the decoupled meson field equation
is parameterized by 
\begin{equation}
\label{fe}
\partial_{\mu}\partial^{\mu}\Phi+m^2\Phi+\lambda_1\Phi^{2p+1}+
\lambda_2
\Phi^{4p+1}=0,
\end{equation}
where $\Phi$ is the operator to describe the self--interacting
fields. For mesons with nonzero spin this operator is a vector in
Minkowski space. 
The parameter $p$ equals $1/2$ or $1$ to yield odd or even powered 
nonlinearities. Using this parameterization, 
equation (\ref{fe}) can be solved analytically. 
The solutions are represented as a power
series in $\varphi$ \cite{Burt81}
\begin{equation}
\label{sol}
\Phi=\sum_{n=0}^{\infty}C_n^{1/{2p}}(w)\;b^n\;\varphi^{2pn+1},
\end{equation}
where $\varphi$ is a solution of the free Klein--Gordon equation with
meson mass $m$. 
These special wavelike solutions of (\ref{fe})
are oscillating functions which propagate with constant shape and 
velocity.  
Corresponding to the classical theory of nonlinear waves 
they shall be called {\it solitary meson fields}.  
The coefficients $C^a_n(w)$ are Gegenbauer Polynomials, 
$b$ and $w$ are
functions of the coupling constants and the order $p$ of the
self--interaction. To quantize the solitary fields we use free 
wave solutions of the Klein--Gordon equation in a finite volume $V$
\cite{fth} 
\begin{equation}
\label{fs}
\varphi(x,k):=\frac{1}{\sqrt{2D_k\omega_kV}}\;a(k)\;
e^{-ikx},
\end{equation}
where the operator $a(k)$ annihilates and the hermitian adjoint
$a^{\dagger}(k)$ creates quanta of positive energy
$$
\omega_k^2=\vec{k}\,^2+m^2.
$$ 
At this point it is important to notice that we added a factor  
$D_k$ which is an
arbitrary Lorentz invariant function of $\omega_k$. As will become 
obvious later, this constant is
crucial for the proper normalization of solitary waves. 
  
The
probability for the propagation of an interacting field can now 
be defined as the amplitude to create an interacting system
at some space--time point $x$ which is annihilated into the vacuum
 at
$y$. Since the intermediate state is not observable and the 
particles
are not distinguishable a weighted sum over all intermediate 
states has to
be performed \cite{Jae95}
\begin{eqnarray}
\lefteqn{iP(y-x) = }\nonumber\\
 & & \sum_k\sum_{N=0}^{\infty}\displaystyle{\frac{1}{N!}}
\Big[\langle
0|\Phi(y,k)|N_k\rangle \langle
N_k|\Phi^{\dagger}(x,k)|0\rangle\theta(y_0-x_0)\nonumber\\
 & & {}+\langle 0|\Phi(x,k)|N_k\rangle \langle
N_k|\Phi^{\dagger}(y,k)|0\rangle\theta(x_0-y_0)\Big].
\label{msmp}
\end{eqnarray}
A straight forward calculation yields the desired amplitude in
momentum space
\begin{eqnarray}
\lefteqn{iP(k^2,m)=\sum_{n=0}^{\infty}\Big[C_n^{1/{2p}}(w)\Big]^2}
\nonumber\\ & & \times\frac{b^{2n}}{(2V)^{2pn}}\frac{(2pn+1)
^{2pn-2}}
{D_k^{2pn+1}(\vec{k}\,^2+M_n^2)^{pn}}\;i\Delta_F(k^2,M_n),
\label{pprop1}
\end{eqnarray}
with the Feynman propagator
\begin{equation}
\label{fprop}
i\Delta_F(k^2,M_n)=\frac{i}{k_{\mu}k^{\mu}-M_n^2},
\end{equation}
and a mass--spectrum
$$
M_n=(2pn+1)m.
$$
Since $V\cdot\omega_k $ is a Lorentz--scalar the amplitude 
(\ref{pprop1}) is
Lorentz invariant. At this point it is convenient to introduce 
the  
dimensionless coupling constants $\alpha$, 
$\alpha_1$ and $\alpha_2$ which we
define as
\begin{eqnarray}
\alpha & := & \displaystyle{\frac{b}{(2mV)^p}}, \nonumber\\
 & & \nonumber\\
\alpha_1 & := & \frac{\lambda_1}{4(p+1)m^2(2mV)^p}, \nonumber\\
 & & \nonumber\\
\alpha_2 & := & \frac{\lambda_2}{4(2p+1)m^2(2mV)^{2p}}.
\label{a1a2}
\end{eqnarray}
This yields
\begin{equation}
\label{wa}
w=\frac{\alpha_1}{\sqrt{\alpha_1^2-\alpha_2}},
\end{equation}
and
\begin{eqnarray}
\label{msprop}
\lefteqn{iP(k^2,m)=\sum_{n=0}^{\infty}\Big[C_n^{1/{2p}}(w)\Big]^2} 
\label{pprop}\\ & & \times\frac{\left[(m^p\alpha_1)^2-m^{2p}
\alpha_2\right]^n 
(2pn+1)^{2pn-2}}{D_k^{2pn+1}(\vec{k}\,^2+M_n^2)^{pn}}\;
i\Delta_F(k^2,M_n).
\nonumber
\end{eqnarray}
The amplitude (\ref{pprop}) shall be referred to as {\it modified 
solitary
meson propagator}. For $p=1$ one gets the amplitude for the 
propagation
of pseudoscalar fields, $p=1/2$ describes scalar particles. 
The series (\ref{msprop}) converges rapidly,  
depending on the mass the subsequent terms diminish by two
($\pi$) or three ($\omega$) orders of magnitude and 
in practical calculations it is sufficient to use 
$n_{max}=4$.
\subsection{Proper Normalization}
The propagator (\ref{pprop}) contains the arbitrary constant 
$D_k$ which can depend on the energy $\omega_k$ and the
coupling constants and is fixed by the conditions
\cite{Burt81}
\begin{itemize}
\item all amplitudes must be Lorentz invariant,
\item $D_k$ must be dimensionless,
\item all self--scattering diagrams must be finite,
\item the fields have to vanish for $\lambda_1,
\lambda_2\to 0$. 
\end{itemize}
Whereas the first three conditions are evident the last 
one requires to recall:
\begin{quote}
If a particle has no interaction then there is no way to
create or measure it and the amplitude for such a process
vanishes. The field exists solely because of its 
interaction.
\end{quote}
The proper normalization constant is a powerful
tool to avoid the problem of regularization which arises in 
conventional 
models. In a $\lambda\Phi^4$--theory for example, which is
described by setting $\lambda_2=0$ and $p=1$, one gets infinite 
results
calculating the first correction to the two--point function
$iP(k^2,m)$. A proper normalization, i.\,e.\ using the smallest power
$$
D_k\sim(\omega_k  V)^2,
$$
makes the result finite. A different situation occurs in models 
including massive spin--1 bosons. Such a case, with or without
self--interaction, is harder to regularize due to the additional 
momentum dependence which arises from the tensor structure in Minkowski 
space.  
Nevertheless, the vector mesons
$\rho$ and $\omega$ are important ingredients in every 
boson exchange
model. A minimum power proper normalization to solve this
problem is 
$$
D_k\sim(\omega_k  V)^4.
$$
In summary, we satisfy the above stated four conditions with
\begin{equation}
\label{dkl}
D_k=\left\{1+
\left[\left(\displaystyle{\frac{m^2}{\lambda_1}}\right)^{\frac{2}{p}}+
\left(\displaystyle{\frac{m^2}{\lambda_2}}\right)^{\frac{1}{p}}\right]
(\omega_kV)^2
\right\}^{\kappa}
\end{equation}
$$
\mbox{where:}\quad\left\{\begin{array}{ll} \kappa
=1 &\quad\mbox{for scalar and}\\
 & \quad\mbox{pseudoscalar mesons.} \\
 & \\
\kappa=2&\quad\mbox{for vector mesons.}
\end{array}\right.
$$
\subsection{The Benefits of OSBEP}
\label{tsl}
With its proper normalization, the solitary meson propagator now is
completely determined and can be applied in a boson exchange
potential. In conventional models a renormalized Feynman propagator is
used to describe the meson propagation. Additionally, an empirical form
factor is attached to each vertex to achieve convergence in the
scattering equation. In a model with solitary mesons as excess particles
the solitary meson propagator (\ref{pprop1}) should be used instead of the
Feynman propagator. Due to the proper normalization, the solitary meson
propagator already carries a strong decay with increasing momentum and
thus offers the possibility to forgo the form factors. Therefore, proper
normalization not only cures the problem to regularize the meson
self--energy but simultaneously provides a meson propagator which is
able to substitute the form factors.
 
Additionally, comparing the properly normalized solitary meson propagators
to the Bonn--B form factors, we find an empirical scaling relation for
the meson self--interaction coupling constants (\ref{a1a2}). We regarded
the simple case $\lambda_2=0$ to obtain \cite{jae96}
\begin{equation}
\label{scaling}
\begin{array}{rrcll}
 & \alpha(m) & = &
\alpha_{\pi}\cdot
\left(\displaystyle{\frac{m_{\pi}}{m}}\right)^{\frac{1}{2}}
 & \quad\mbox{for scalar fields,}\\
 & & & & \\
\mbox{and:}\quad & \displaystyle{\frac{\alpha(m)}{\sqrt{\kappa}}} 
& = &
\displaystyle{\frac{\alpha_{\pi}}{\sqrt{\kappa_{\pi}}}}
\cdot\left(\displaystyle{\frac{m_{\pi}}{m}}\right)
 & \quad\mbox{for pseudoscalar} \\
 & & & & \quad\mbox{and vector fields.}
\end{array}
\end{equation}
Thus, the pion self--interaction coupling constant $\alpha_{\pi}$ is the
only parameter to describe the meson dynamics. This connection between
masses and coupling constants can be anticipated in a model which is
motivated by QCD and thus puts physical significance into the
parameter $\alpha_{\pi}$. 
  
\begin{figure}[t] \centering
\begin{picture}(7.0,6.125)(0.0,0.0)
\epsfig{figure=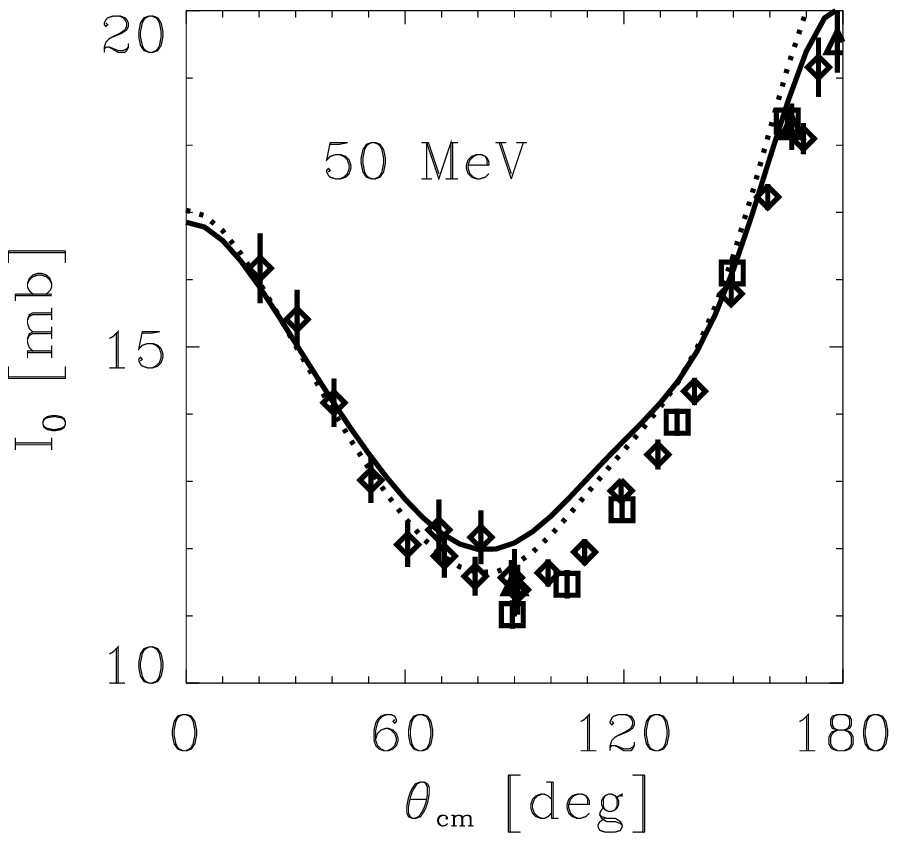,width=7.0cm}
\end{picture}
\begin{picture}(7.0,6.125)(0.0,0.0)
\epsfig{figure=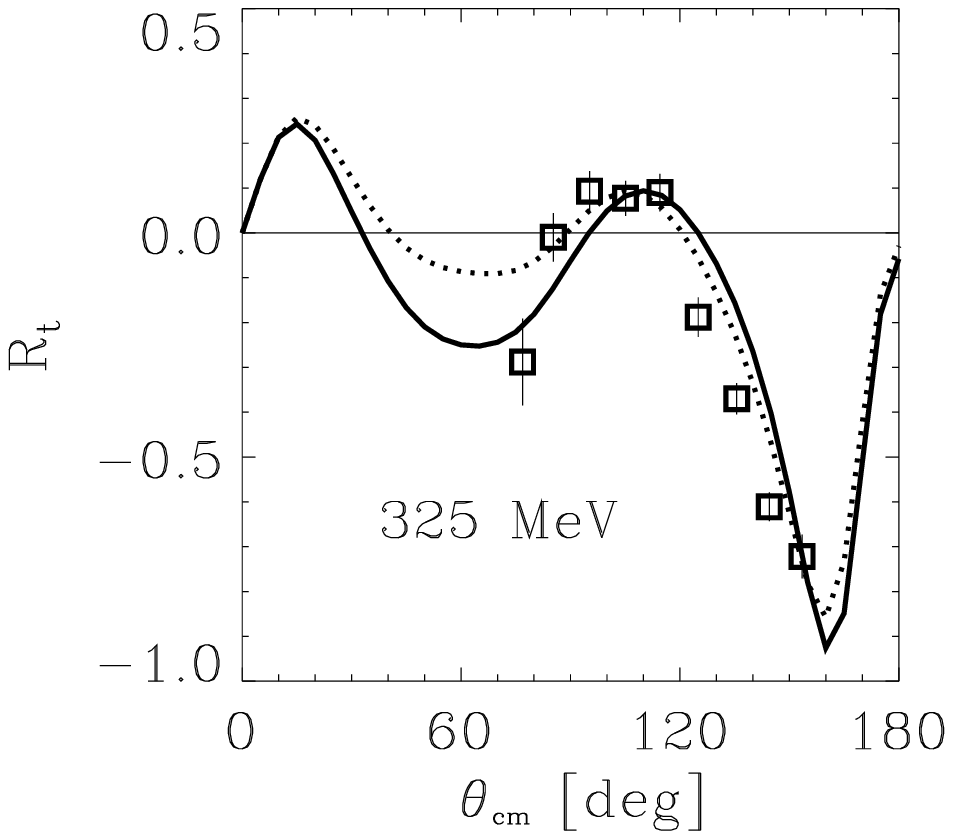,width=7.0cm}
\end{picture}
\begin{picture}(7.0,6.125)(0.0,0.0)
\epsfig{figure=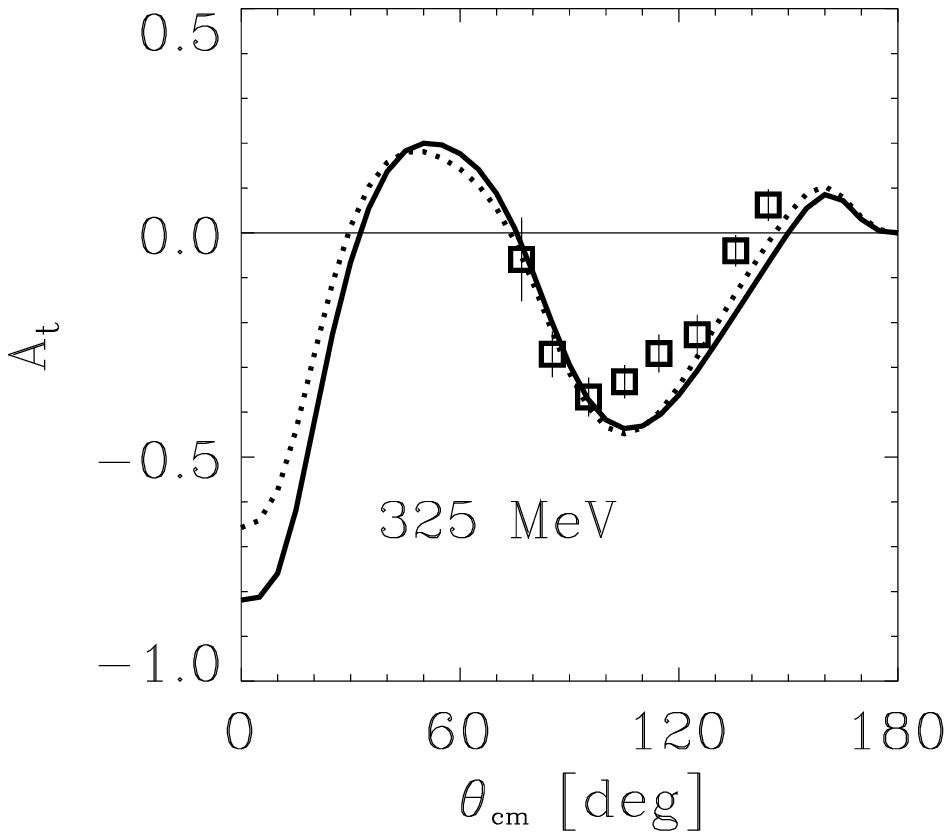,width=7.0cm}
\end{picture}
\begin{picture}(7.0,6.125)(0.0,0.0)
\epsfig{figure=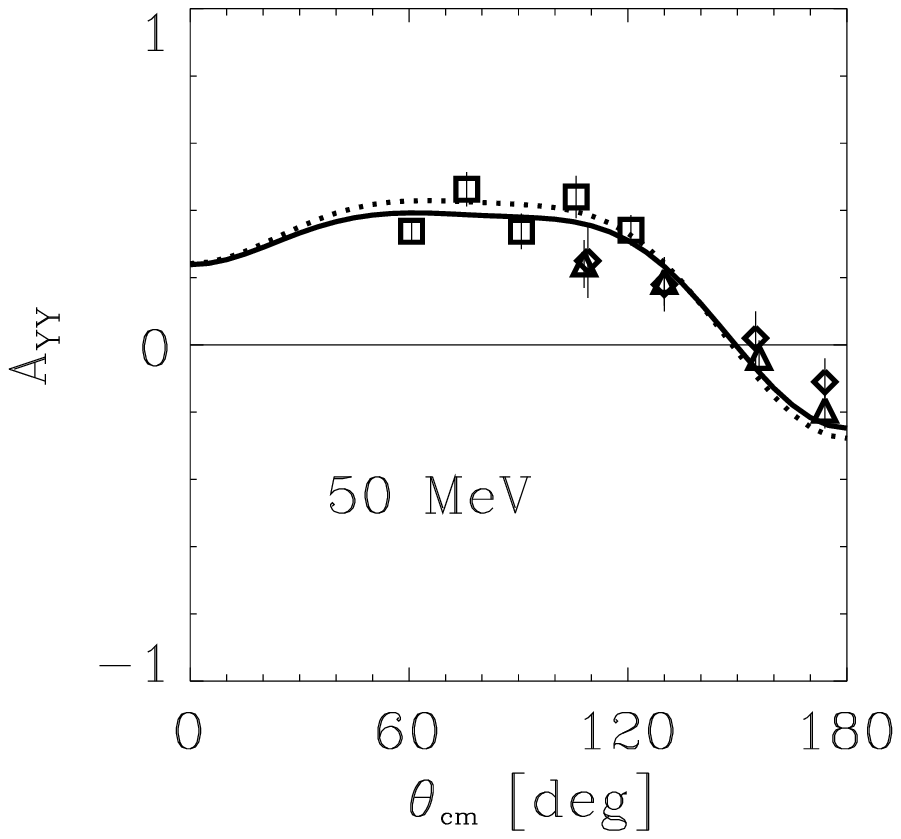,width=7.0cm}
\end{picture}
\begin{picture}(7.0,6.125)(0.0,0.0)
\epsfig{figure=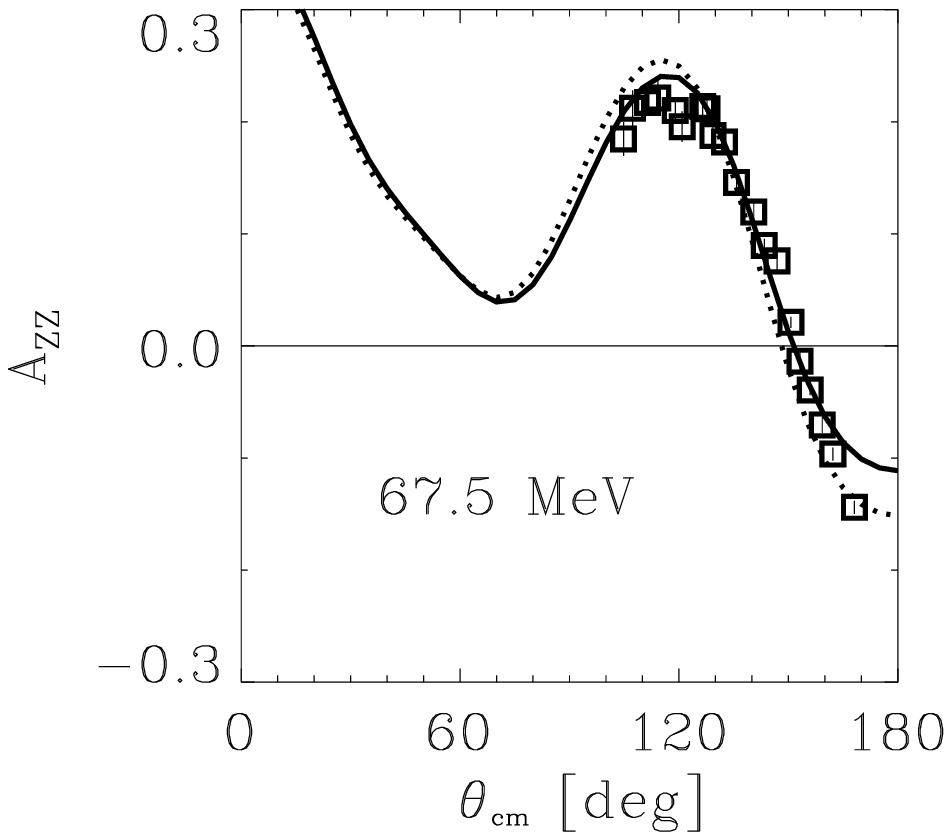,width=7.0cm}
\end{picture}
\begin{picture}(7.0,6.125)(0.0,0.0)
\epsfig{figure=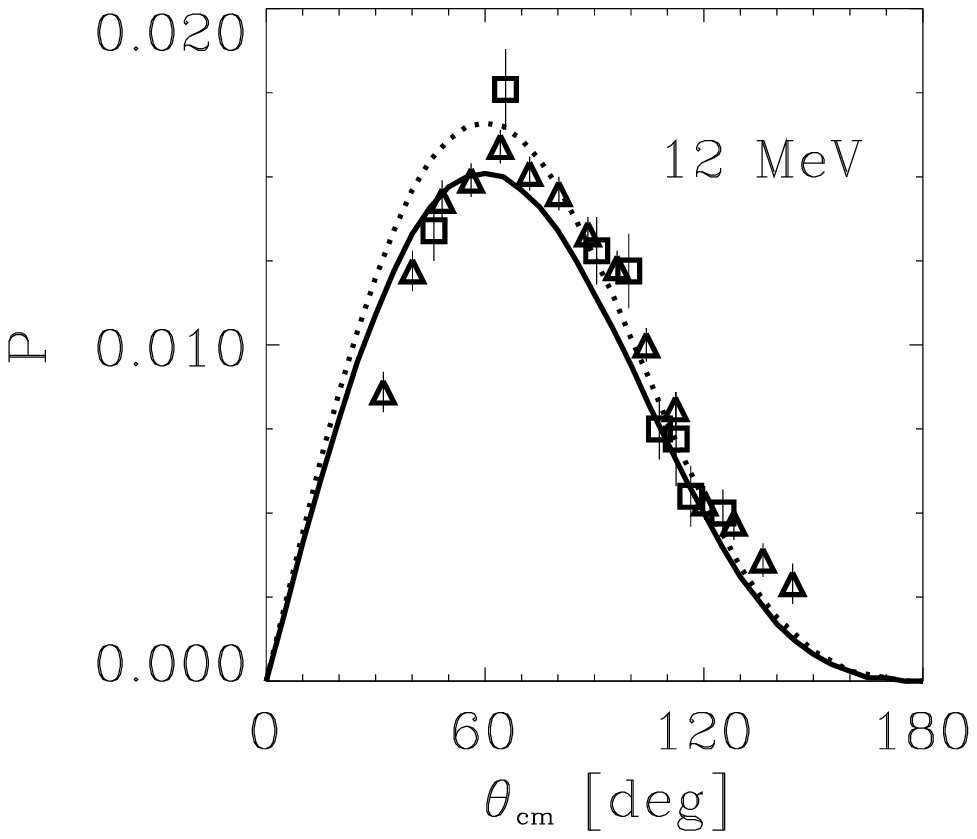,width=7.0cm}
\end{picture}
\caption[Observables of $np$ scattering]
{Observables of $np$ scattering. 
Kinetic laboratory energy is denoted, experimental data are taken from
the VPI--SAID program. We show theoretical predictions from OSBEP
(full) and Bonn--B (dotted). \label{obs}}
\end{figure}

\begin{figure}[b] \centering
\begin{picture}(7.5,5.16)(0.0,0.0)
\epsfig{figure=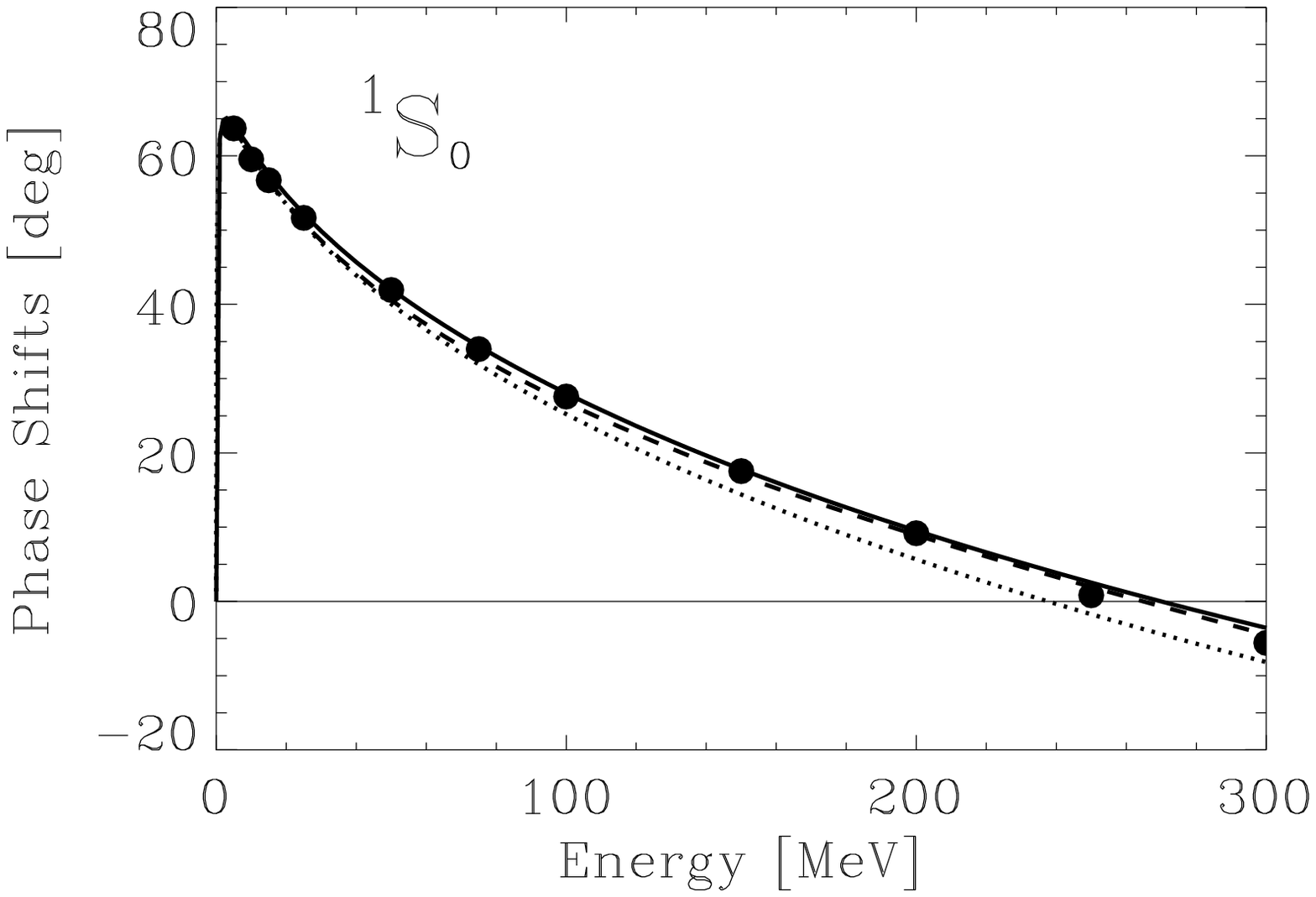,width=7.5cm}
\end{picture}
\begin{picture}(7.5,5.16)(0.0,0.0)
\epsfig{figure=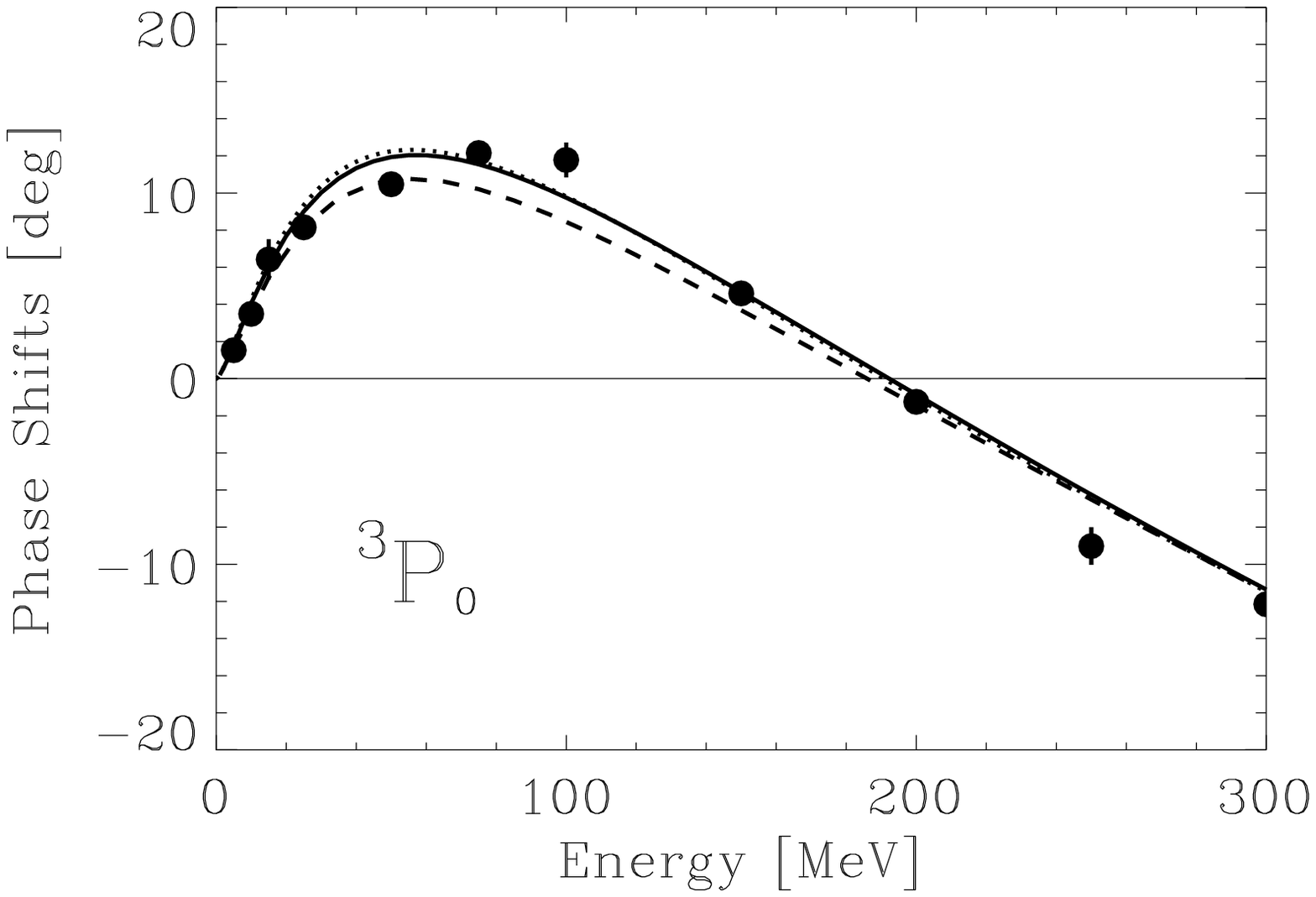,width=7.5cm}
\end{picture}
\\
\begin{picture}(7.5,5.16)(0.0,0.0)
\epsfig{figure=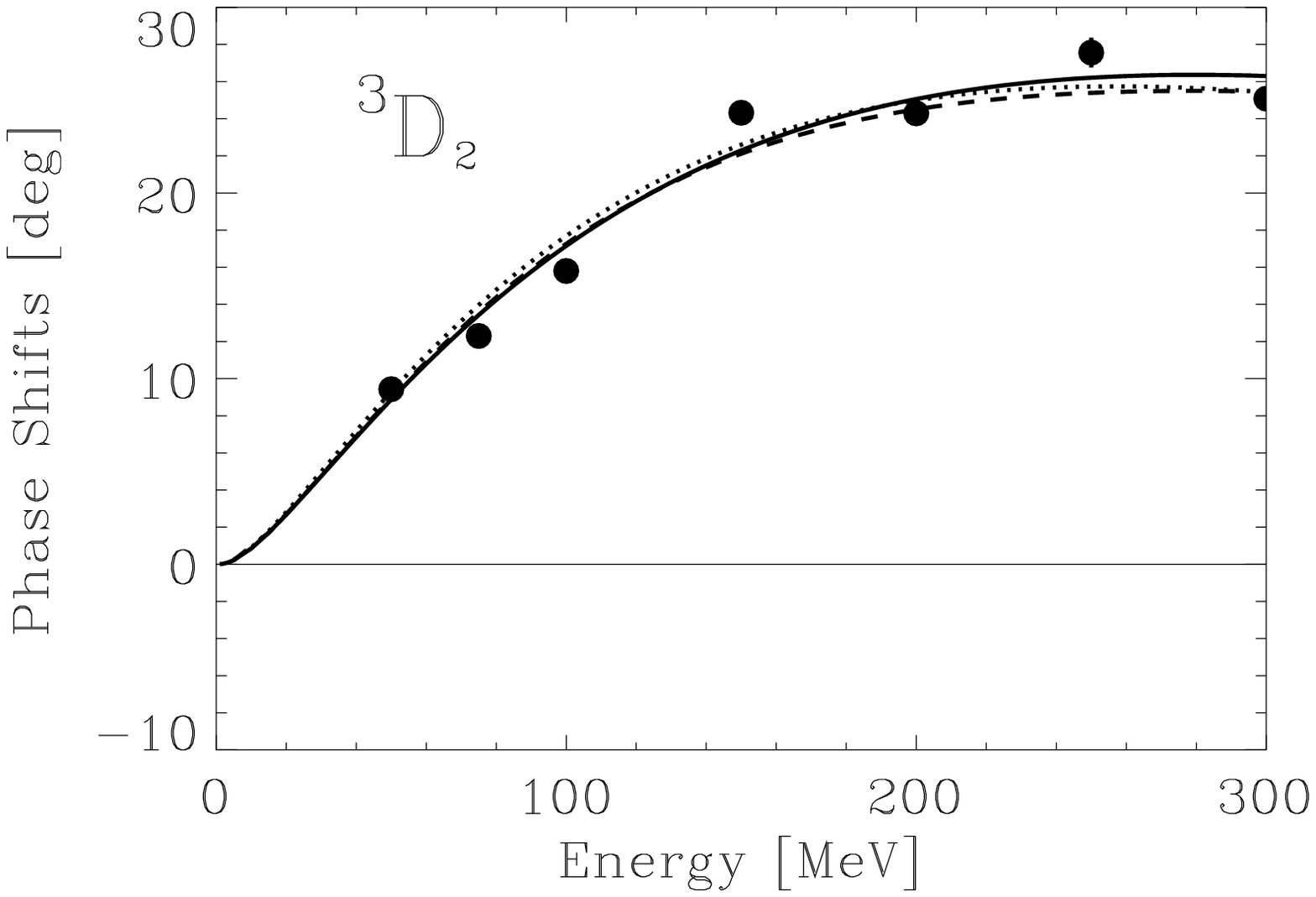,width=7.5cm}
\end{picture}
\begin{picture}(7.5,5.16)(0.0,0.0)
\epsfig{figure=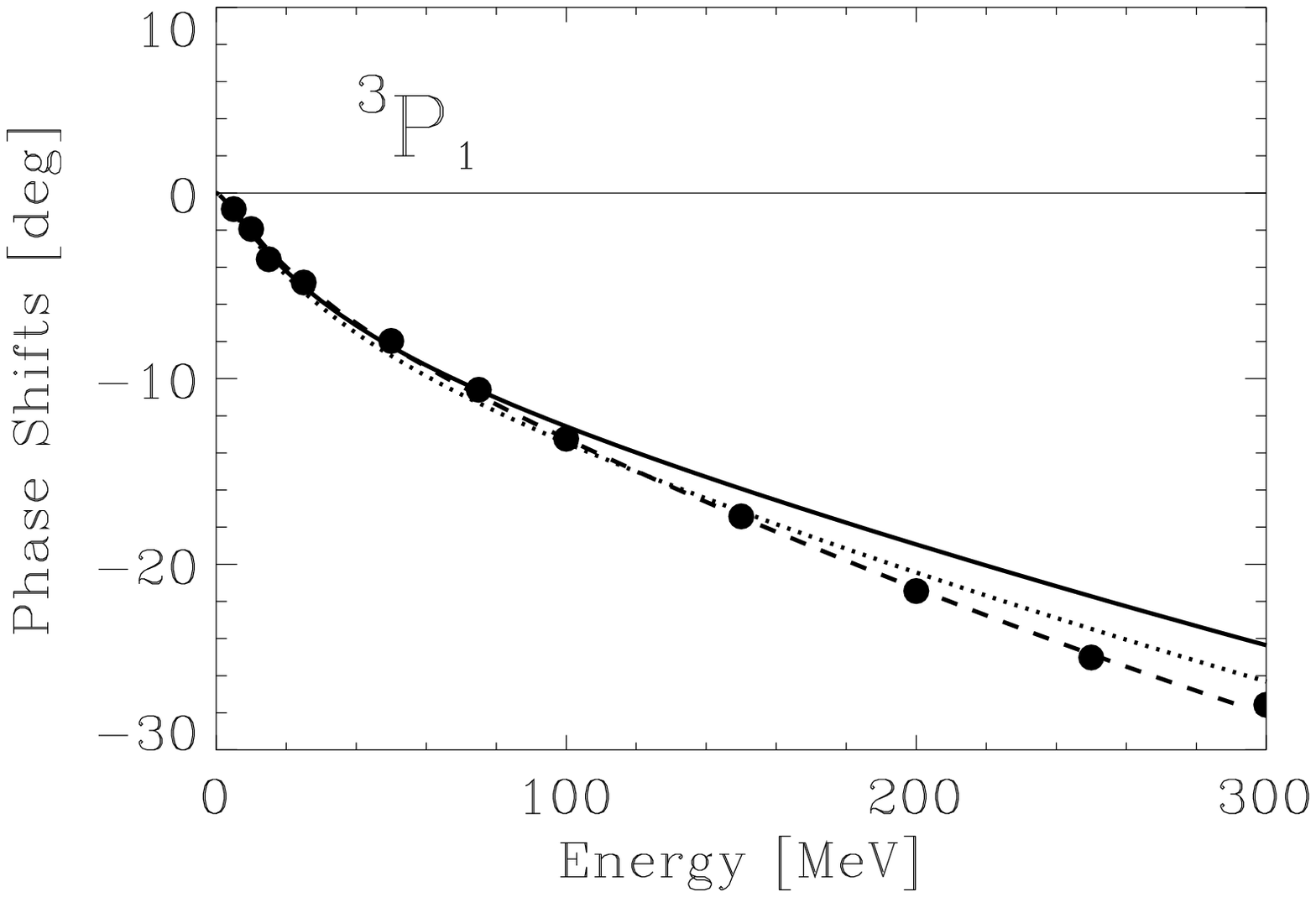,width=7.5cm}
\end{picture}
\\
\begin{picture}(7.5,5.16)(0.0,0.0)
\epsfig{figure=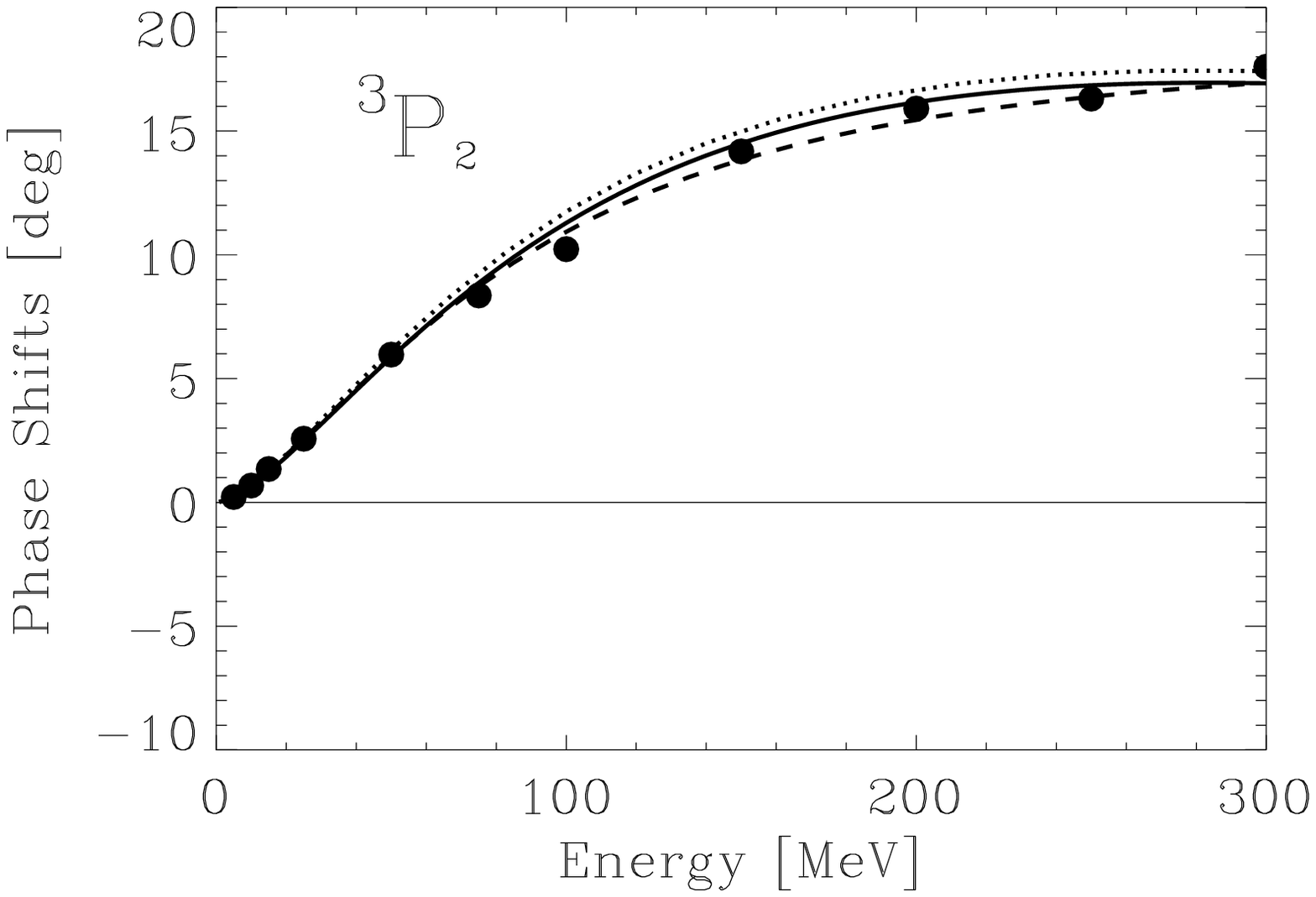,width=7.5cm}
\end{picture}
\begin{picture}(7.5,5.16)(0.0,0.0)
\epsfig{figure=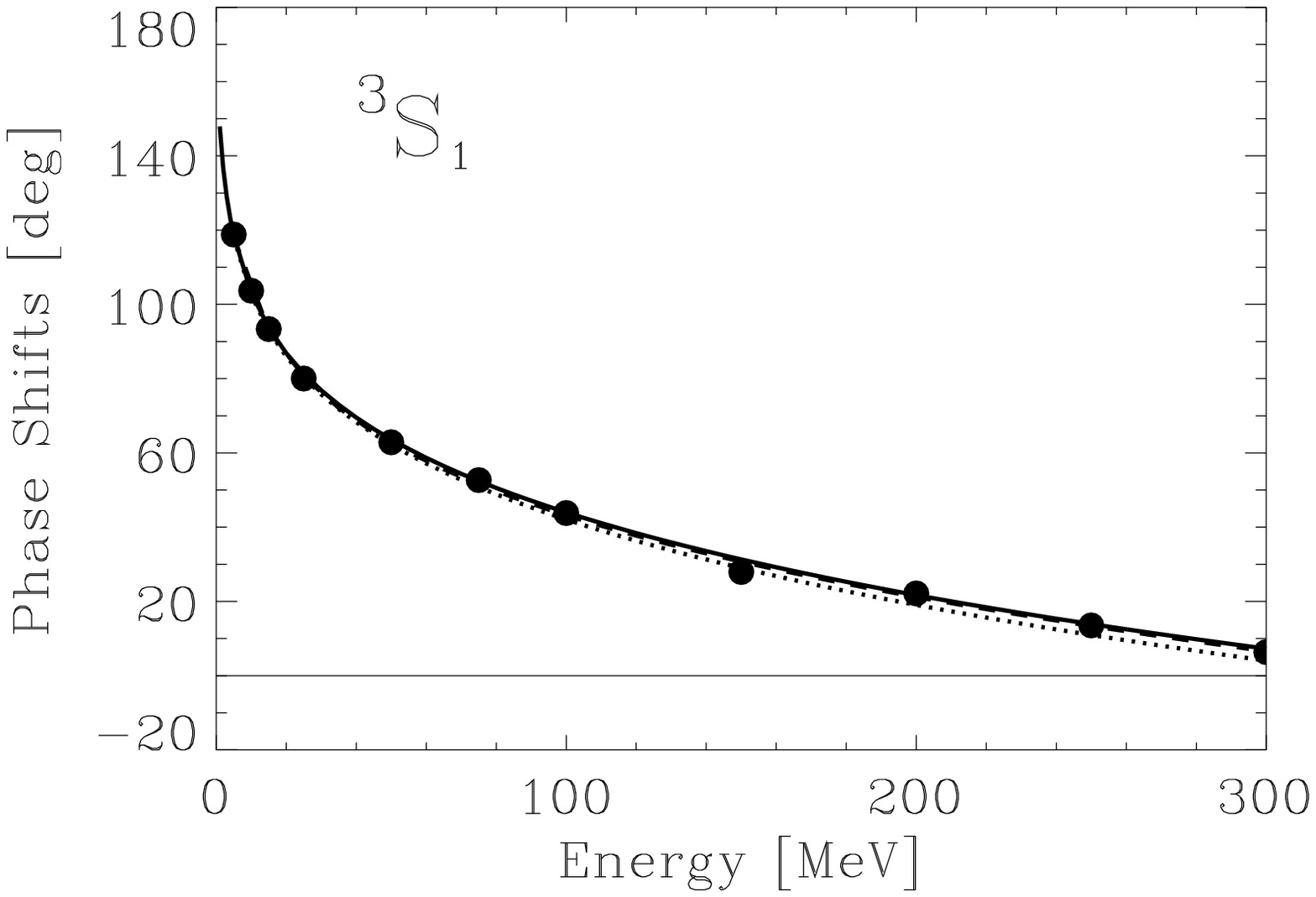,width=7.5cm}
\end{picture}
\caption[SYM $^3SD_1$ Phase Shifts]
{\small Selected $np$ phase shifts: Arndt SM95 (circles),
Nijmegen PWA (dashed), Bonn--B (dotted) and OSBEP (full).
\label{phsh}}
\end{figure}

Together with the
meson--nucleon coupling constants the model contains a total number of
nine parameters (see Table\,\ref{coco}) which are adjusted 
to obtain a good fit to $np$ scattering
phase shifts and deuteron data (see Table\,\ref{deut} and
Figure\,\ref{phsh}). The agreement with scattering observables 
(Figure\,\ref{obs}) is excellent and the goal
to achieve a description of experimental data which is
comparable to the Bonn--B potential is accomplished.  
\section{The quest for Off--Shell Effects}
Concerning nucleon--nucleon scattering, boson exchange models and
inversion potentials show excellent agreement with experimental
data. The results of the OSBEP and Bonn--B models are shown in
Figure\,\ref{phsh}, the inversion potentials reproduce the phase shifts
by construction. The next step is to apply both potential models in more
complex reactions to find out whether the absent off--shell
information in the inversion potentials would lead to a break--down of
the model in situations where the off--shell part of the scattering
amplitude contributes. Three of such processes have been considered and
shall be reviewed here: Application of model potentials in ($p$,
$p\gamma$) Bremsstrahlung \cite{ppg}, calculations of triton binding
energy \cite{trit} and --- most recently --- usage of boson exchange
and inversion potentials in an in--medium, full--folding optical
potential model for nucleon--nucleus scattering \cite{na}. 
  
The strategy to compare
boson exchange and inversion potentials is as follows. We take some
model potential $t$--matrix $t_{mod}(k,k^{'})$ with its characteristic
off--shell (i.\,e.\ $k\neq k^{'}$) behavior, calculate the phase shifts
from the on--shell part $t_{mod}(k,k)$, use these model phase shifts
to calculate an inversion potential $V_{\ell}(r)$, as described in
Section\,\ref{inv}, which in turn yields a $t$--matrix
$t_{inv}(k,k^{'})$ where the
off--shell behavior is uniquely defined by the restriction \cite{inv}
\begin{equation}
\int_a^{\infty}r|V_{\ell}(r)|\;dr\;<\;\infty\qquad\mbox{for}\qquad a\geq0.
\end{equation}
Therefore, any on--shell difference of inversion potentials will lead to
off--shell differences in the $t$--matrix. On the other hand, the
off--shell behavior of the model $t$--matrix and the $t$--matrix
obtained from inversion will in general be entirely
different, at least for large momenta. Therefore, comparing
$t_{mod}(k,k^{'})$ and $t_{inv}(k,k^{'})$ in off--shell sensitive
reactions works as a tool to test whether off--shell differences of
on--shell equivalent $t$--matrices manifest themselves in the
calculation of observable data. 
\subsection{Cross Sections for ($p$, $p\gamma$) Bremsstrahlung}
\begin{figure}[b] \centering
\begin{picture}(7.64,6.36)(0.0,0.0)
\epsfig{figure=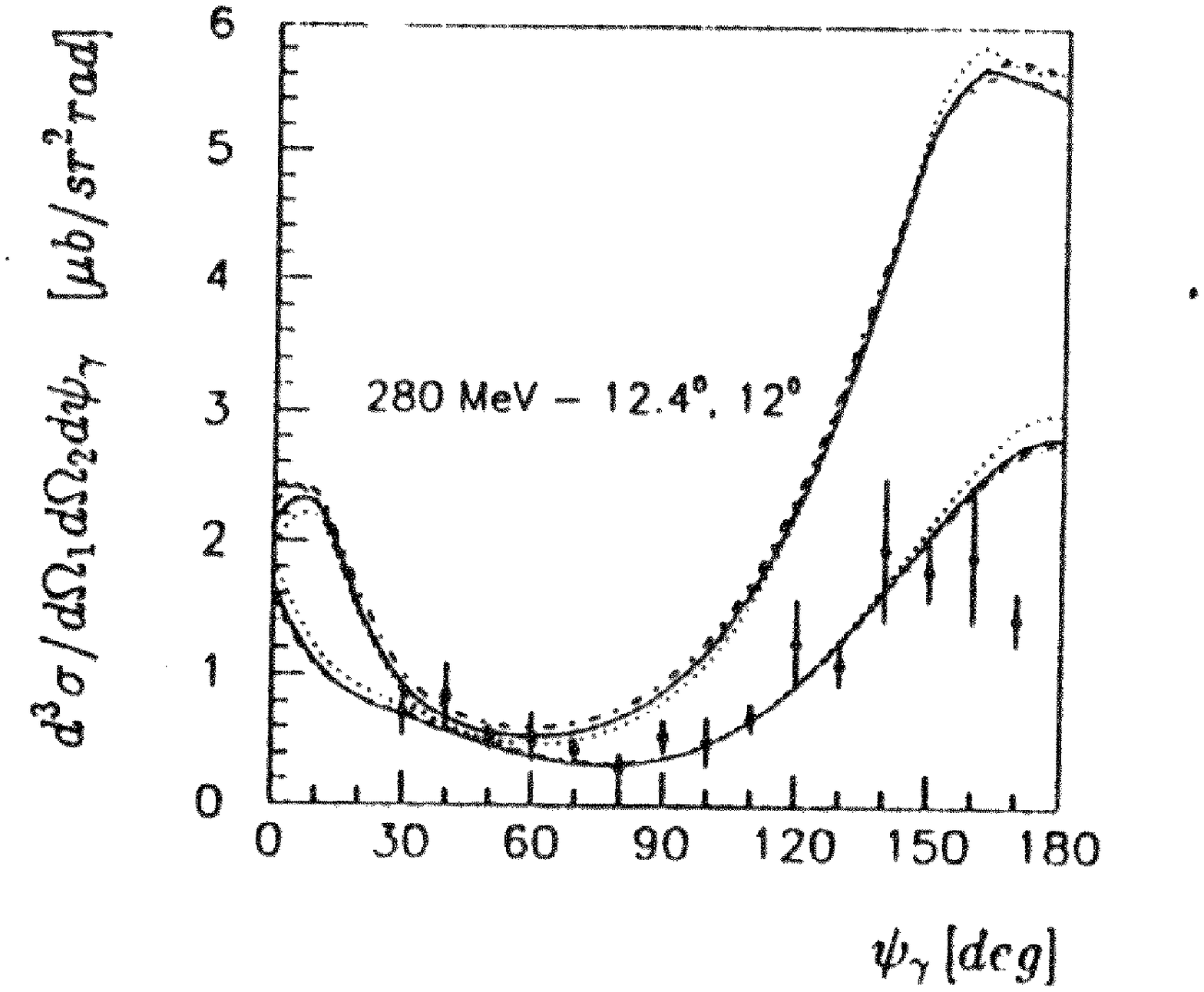,width=7.64cm}
\end{picture}
\begin{picture}(6.36,4.36)(0.0,0.0)
\epsfig{figure=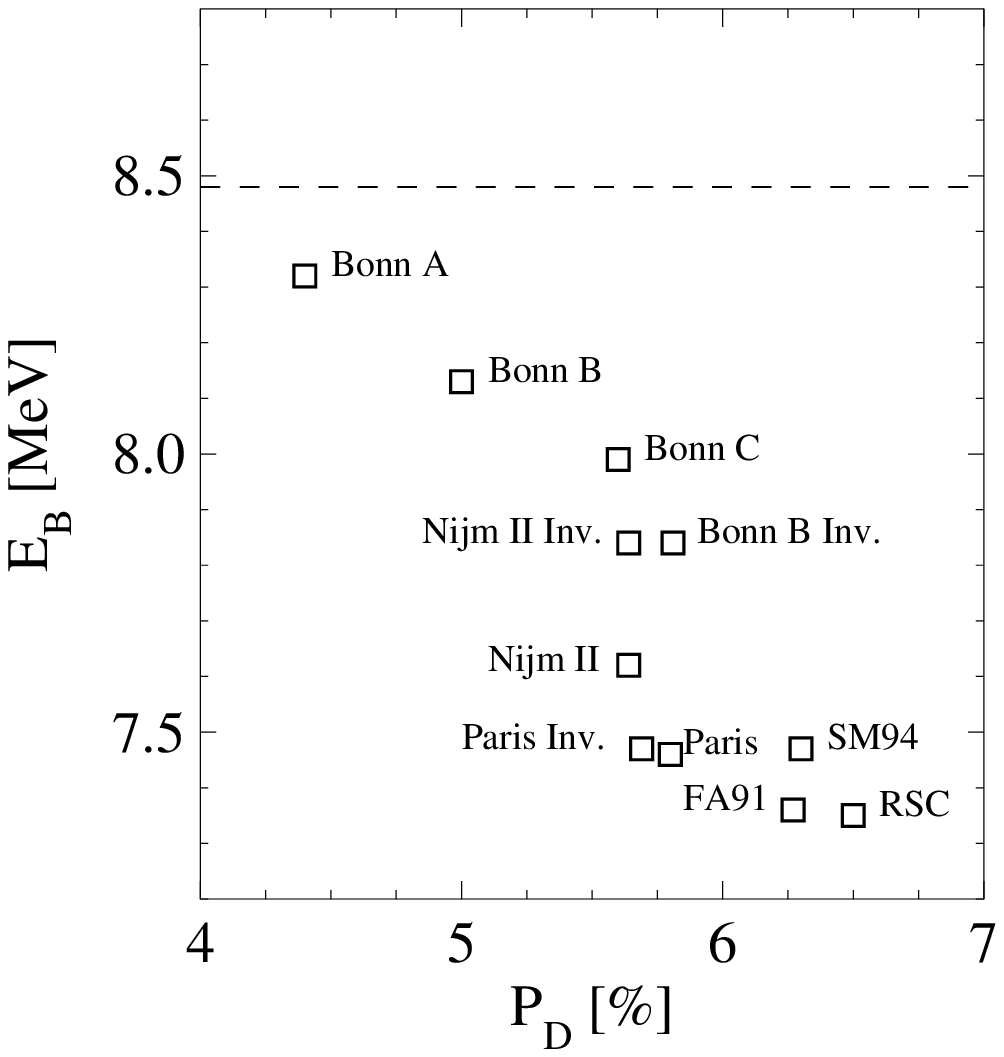,width=6.36cm}
\end{picture}
\\
\caption[]
{\small Left figure: Theoretical ($p$, $p\gamma$) 
cross sections from various
boson exchange and inversion potentials. The 
lower curves show the
results for an exact calculation whereas the 
upper curves represent an
on--shell approximation. The lines are Nijmegen 
PWA inversion (solid), Paris
(dash--dotted), Bonn--B (dashed) and an inversion potential from Bonn--B
phase shifts (dotted). Right figure: Triton binding energy versus deuteron
$D$--state probability for various potentials. Note the large differences
in both entities for the genuine and inversion Bonn--B potential,
whereas Paris original and Paris inversion 
show similar results. \label{bild2}}
\end{figure}

In Figure\,\ref{bild2}, the theoretical ($p$, $p\gamma$) cross sections
for several potentials are shown. In detail, Jetter et al.\ \cite{ppg}
calculated cross sections using the boson exchange potentials Paris and
Bonn--B \cite{pot} as well as inversion potentials from the
Nijmegen PWA \cite{nijm} and phase shifts from the Bonn--B
potential. Two kind of calculations for ($p$, $p\gamma$) cross sections 
were applied. First, an
on--shell approximation which is shown as upper curve in
Figure\,\ref{bild2} and clearly misses the data, second an exact
off--shell calculation which is represented by the lower
curve. Obviously, using the exact calculation 
all potentials are consistent with the
data, boson exchange models leading to equivalent results as
inversion potentials. 
Additionally, as a surprising result the cross section for ($p$,
$p\gamma$) Bremsstrahlung, 
which was assumed to be sensitive to the off--shell behavior
of the $t$--matrix, can not distinguish between the original Bonn--B
potential (dashed line) and the on--shell equivalent but off--shell
different inversion potential (dotted) 
obtained from Bonn--B model phase shifts!  
Consequently, nothing can be learned about the off--shell properties 
of the NN scattering amplitude from ($p$,
$p\gamma$) Bremsstrahlung.
\subsection{Triton Binding Energies}
Analogue to ($p$, $p\gamma$) Bremsstrahlung calculations, a number of
potentials were applied to compute binding energies of $^3$H
\cite{trit}. As representative examples we consider  
model and inversion potentials for the Paris and Bonn--B
potentials. The results follow the well--known correlation between the 
deuteron $D$--state probability and the binding energy $E_B(^3\mbox{H})$
which is shown in Figure\,\ref{bild2} for various 
models. All of the model predictions underestimate the experimental
value of 8.48\,MeV, a circumstance which will be discussed below. 
For the Bonn--B and Paris potential the exact values are listed in
Table\,\ref{pdtrit}. 
Note that for the Paris original and inversion 
potential the triton binding energies are the
same whereas for Bonn--B a significant difference arises. The reason can
be seen from an argument pointed out by Machleidt \cite{mach}. The
on--shell $t$--matrix is related to the central and tensor potential by
the approximate relation
\begin{equation}
\label{tma}
t(k,k) \approx V_C(k,k)-\int
d^3\vec{k}^{'}\;V_T(k,k^{'})\frac{M}{k^{'}{^2}-
k^2-i\epsilon}V_T(k^{'},k).
\end{equation}
Thus, for the genuine Bonn--B and the Bonn--B 
inversion potential, which are
on--shell equivalent, the {\it sum} of Born term and integral
term in (\ref{tma}) is equal, while the respective 
terms themselves may be
different. The deuteron $D$--state probability 
on the other hand is
dominantly determined by the tensor potential 
\begin{equation}
\label{pd}
P_D\approx
\int_0^{\infty}\!\int_0^{\infty}k^2dk
\;{k^{'}}{^2}dk^{'}\;\frac{V_T(k,k^{'})}
{-E_B(^2\mbox{H})-\frac{k^2}{M}}\;\Psi_0(k),
\end{equation}
where $\Psi_0(q)$ is the deuteron $S$--wave. Thus, 
two on--shell
equivalent potentials may indeed lead to different 
$D$--state
probabilities and triton binding
energies due to differences in 
the tensor potential parts. Obviously, this is the
case for the Bonn--B original and inversion potential. This can be
understood from the fact that inversion potentials are first of all
local functions of $r$ in coordinate space since the scattering phase
shifts as functions of $k$ build a one--dimensional manifold (see
Section\,\ref{inv}). However, the microscopic
boson exchange potentials, which are naturally formulated in momentum
space, are not necessarily local and this deviation may account for the
differences in the tensor potential.  
Comparing the results for the Bonn--B and Paris potentials, one
finds that the non--locality of the Paris potential obviously can be
well represented by an on--shell equivalent local potential, whereas the
non--locality of the Bonn--B potential produces a significant decrement
of the $D$--state probability. This arises from the fact that the tensor
force in the Paris potential is local, whereas the tensor potential in
the Bonn--B potential is not. Thus, the nonlocality of the other
potential contributions essentially play no role in the calculation of
the triton binding energy.  
  
After all, the calculation of triton binding energies shows some
sensitivity to the non--local structure of potentials. Unfortunately,
there is an ongoing discussion about the influence of three--body forces
and relativistic corrections on the triton binding energy 
so that none of the above results can be favored. Further work on this
field seems desirable. 
\subsection{Nucleon--Nucleus Scattering}
\begin{figure}[t] \centering
\begin{picture}(7,4.33)(0.0,0.0)
\epsfig{figure=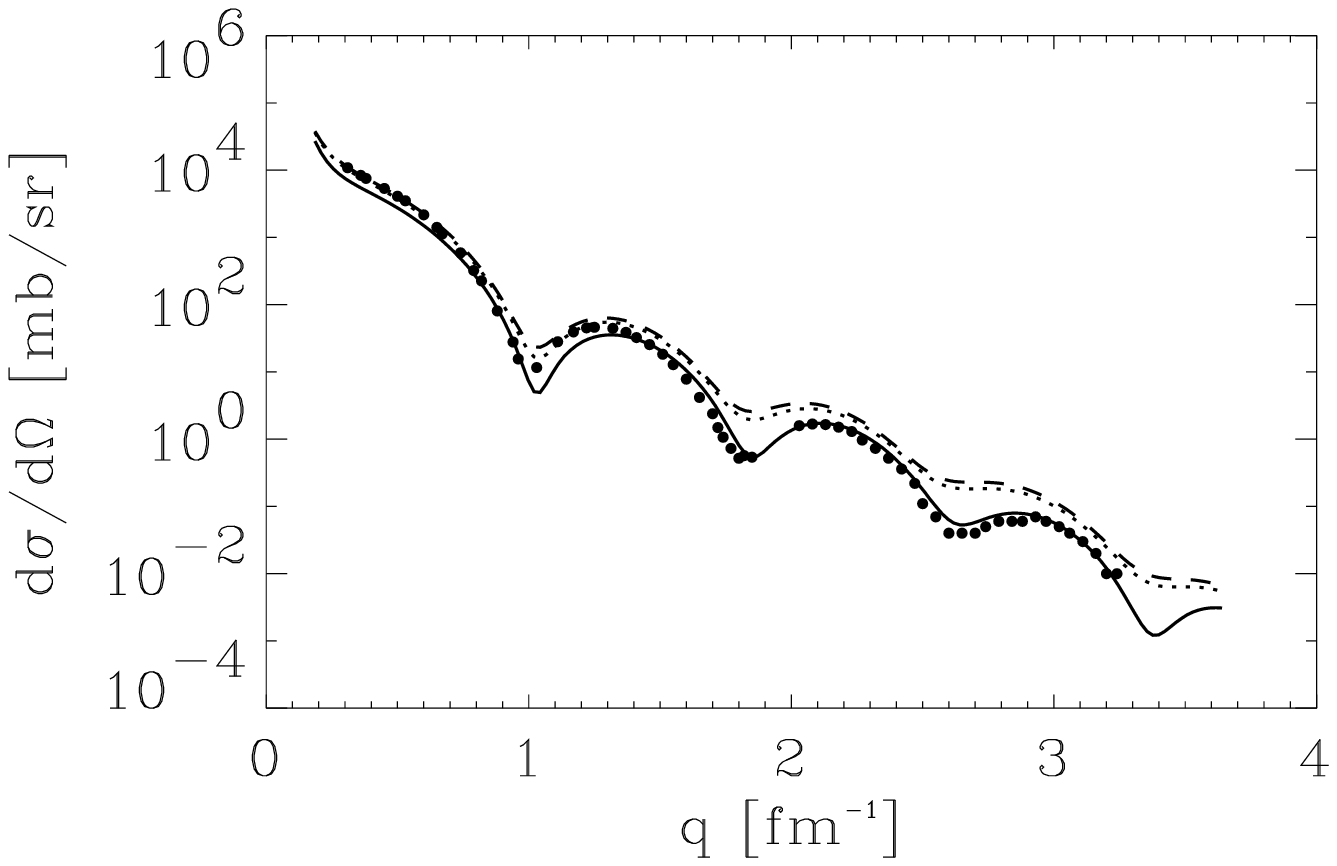,width=7cm}
\end{picture}
\begin{picture}(7,4.33)(0.0,0.0)
\epsfig{figure=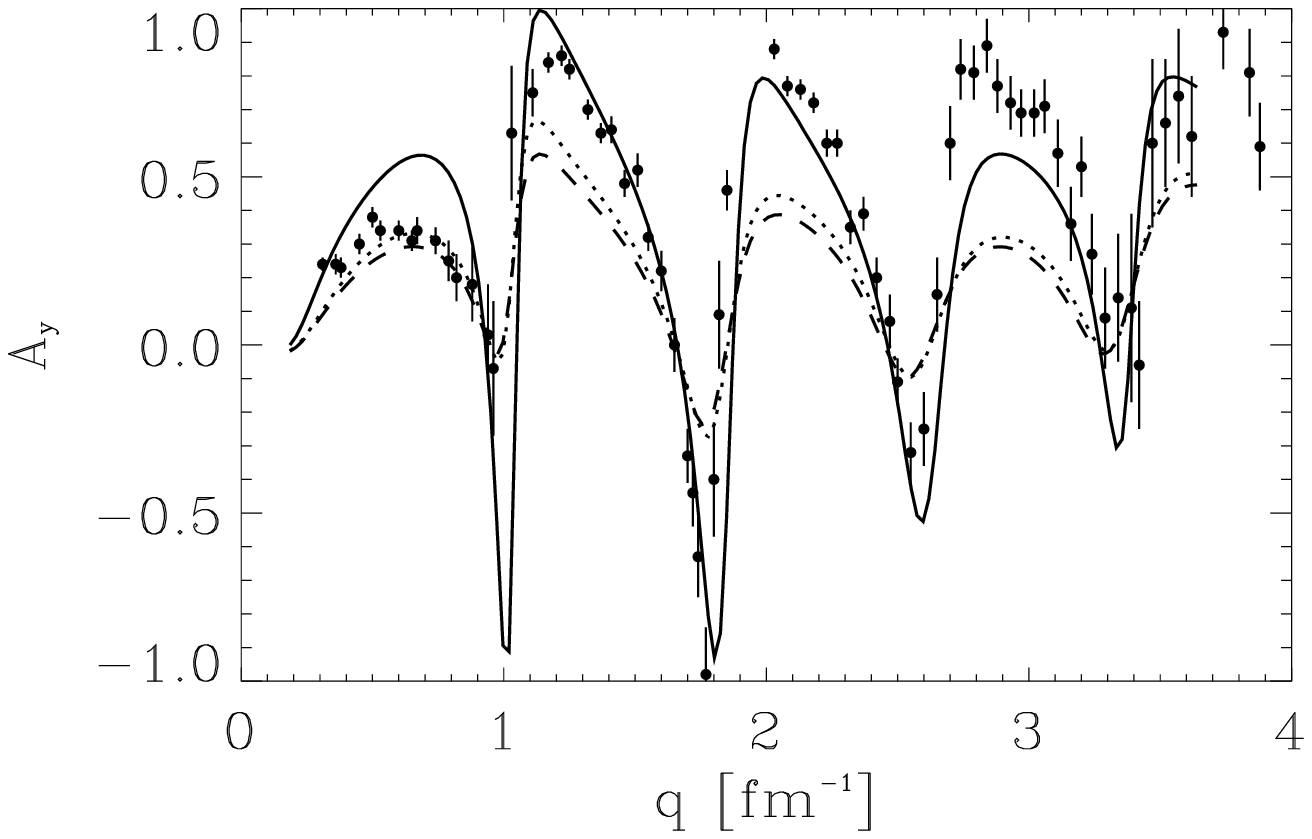,width=7cm}
\end{picture}
\\
\caption[]
{\small Differential cross section (left) and analyzing power (right)
for $^{40}$Ca($p,p$) at 500 MeV. We show results for SM94 inversion
(solid), genuine Paris potential (dotted) and Paris inversion
(dashed). \label{nascat}} 
\end{figure}

Embedded in an in--medium full--folding optical potential model various
boson exchange and inversion potentials have been applied to calculate
observables of nucleon--nucleus scattering \cite{na}. In
Figure\,\ref{nascat} we show the results obtained for the differential
cross section and analyzing power for $^{40}$Ca($p,p$) scattering at
500\,MeV using the Paris potential together with inversion potentials
from Paris model phase shifts and phases from the Arndt SM94 PWA
\cite{arn94}. The conclusions are twofold: As in the case of ($p$,
$p\gamma$) Bremsstrahlung, the genuine Paris and Paris inversion
potential results can not be distinguished by the experiment and thus no
off--shell sensitivity can be found in the analysis of elastic
nucleon--nucleus scattering. Additionally, a new effect occurs comparing
the inversion potentials from SM94 phase shifts and the Paris
potential. The SM94 inversion potential, which by construction fits the
elastic NN data much better than the Paris potential, yields significant
better results for the cross section and the analyzing
power. Conclusively, an improvement in the description of on--shell data
also yields better results in nucleon--nucleus scattering. 
\section{Summary and Outlook}
As described in Section\,\ref{inv} and \ref{osbep}, quantum inversion and
boson exchange potentials rest on entirely different footings. Whereas the
latter are derived from some microscopic model, as the model of solitary
mesons, leading to a momentum space potential with
--- in general --- nonlocal behavior, the inversion potentials are
obtained model--independently from the Gel'fand--Levitan or Marchenko
algorithms. Since elastic NN scattering data enter the inversion
potentials via the phase shifts $\delta_{\ell}(k)$ as a one--dimensional
manifold, inversion potentials are local, energy--independent functions
in coordinate space. This fundamental difference could be expected to
produce significant deviations in the description of processes where the
off--shell part of the $t$--matrix, which does not enter the phase
shifts, contributes. 
 
Surprisingly, the application of boson exchange and inversion
potentials in the calculation of ($p$, $p\gamma$) Bremsstrahlung cross
sections and nucleon--nucleus scattering observables produces equivalent
results for the boson exchange potentials an their local
counterparts. Additionally, in the case of nucleon--nucleus scattering
it turned out that an improvement of the description of on--shell data
also yields better results for off--shell observables.  
  
The only difference in the comparison of boson exchange versus quantum
inversion potentials occurred in the calculation of triton binding
energies. It turned out that the usage of local potentials enhances the
$D$--state probability with respect to their non--local on--shell
equivalents, leading to a lower triton binding energy. This effect could
be traced back to differences in the tensor part of the potential which
can differ for local and non--local potentials. Unfortunately, neither
boson exchange nor quantum inversion potentials can be favored in this
context due to the persistent uncertainty concerning the effect of
three--body forces and relativistic corrections on the triton binding
energy. 
  
To our disappointment, the study of the above reactions thus is not
suitable to put boson exchange potentials to a comparative
test. Therefore, as a future prospect, the application of 
boson exchange and inversion
potentials in meson--nucleon and meson--meson scattering 
seems promising to test the
concept of quantum inversion as well as the applicability of NN
potential models in a wider range of
hadron--hadron interactions. In this context, the OSBEP
model can be expected to show interesting results. In contrast to
conventional boson exchange potentials, which use different form 
factors in nucleon--nucleon and
meson--nucleon interactions, the concept of proper normalization
in OSBEP remains unchanged. 
Thus, the OSBEP model may be able to describe both
interactions consistently with a very small number of parameters. 
\acknowledgements
Supported in part by FZ J\"ulich, COSY Collaboration, Grant Nr.\ 
41126865.
\clearpage
\tighten
\begin{table}  
\caption{OSBEP parameters}
\label{coco}
\begin{tabular}[b]{llllllll}
 & $\pi$ & $\eta$ & $\rho$ & $\omega$ & $\sigma_0$ & $\sigma_1$ 
& $\delta$ \\
\hline
$S^P$ & $0^-$ & $0^-$ & $1^-$ & $1^-$ & $0^+$ & $0^+$ & $0^+$ \\
 $\displaystyle{\frac{g_{\beta}^2}{4\pi}}$ &
    13.7   & 1.3985  & 1.1398 & 18.709 & 14.147 & 7.8389
& 1.3688 \\
\multicolumn{4}{c}{$\alpha_{\pi}=0.428321$} &
\multicolumn{4}{c}{$f_{\rho}/g_{\rho}=4.422$} \\
\end{tabular}
\end{table}
\begin{table} 
\caption{Deuteron Properties}
\label{deut}
\begin{tabular}[t]{llll}
 & \multicolumn{1}{l}{Bonn--B\tablenotemark[1]} &
\multicolumn{1}{l}{OSBEP} & 
\multicolumn{1}{l}{Exp.} \\
\hline
$E_B\mbox{ (MeV)}$ & $2.2246$ &    $2.22459$   
&  $2.22458900(22)$ \\
$ \mu_d$           & $0.8514$\tablenotemark[2]  &  
$0.8532$\tablenotemark[2] &  $0.857406(1)$ \\
$Q_d\mbox{ (fm$^2$)}$ &  $0.2783$\tablenotemark[2] &
$0.2670$\tablenotemark[2] &  $0.2859(3)$   \\
$A_S\mbox{ (fm$^{-1/2}$)}$  &  $0.8860$    &
   $0.8792$    &   $0.8802(20)$  \\
$D/S$           &   $0.0264$    &           
   $0.0256$    &   $0.0256(4)$   \\
$r_{RMS}\mbox{ (fm)}$ &  $1.9688$     &      
  $1.9539$     &  $1.9627(38)$     \\
$P_D\quad(\%)$     &   $4.99$      &          
   $4.6528$      &
\multicolumn{1}{c}{---}   \\
\end{tabular}
\tablenotetext[1]{Data from \protect\cite{Mach89}}
\tablenotetext[2]{Meson exchange current
contributions not included}
\end{table}
\begin{table} 
\caption{Deuteron $D$--State Probabilities and Triton Binding Energies}
\label{pdtrit}
\begin{tabular}[t]{lllll}
 & Paris & Paris (inv.) & Bonn--B &  Bonn--B (inv.) \\\hline
 $P_D$  & 5.77 \% & 5.69 \% & 4.99 \% & 5.81 \% \\
$E_B(^3\mbox{H})$ & 7.47 MeV  & 7.47 MeV  &
8.14 MeV & 7.84 MeV  \\
\end{tabular}
\end{table}
\clearpage
%
%
\clearpage
%
%
%
\clearpage
%
%
%
%
%
%
%

\begin{thebibliography}{23}
%
%
\bibitem{pot}
M.\,M. Nagels, T.\,A. Rijken, and J.\,J. de Swart, Phys.\,Rev.\ D 
{\bf 17}, 768 (1978); 
M. Lacombe, B. Loiseau, J.\,M. Richard, R. Vinh Mau, 
J. C\^{o}t\'{e},
P. Pir\`{e}s, and R. de Tourreil, Phys.\,Rev.\ C {\bf  21}, 861
(1980); 
R. Machleidt, K. Holinde, and Ch.\ Elster, Physics Reports 
{\bf 149}, 
1 (1987). 
%
\bibitem{chi}
T.\,H.\,R. Skyrme, Nucl.\,Phys.\ {\bf 31}, 556 (1962);
S. Weinberg, Phys.\,Rev.\,Lett.\ {\bf 18}, 188 (1967); 
J. Wambach, in {\it Quantum Inversion Theory
and Applications}, edited by H.\,V. von Geramb (Lecture Notes in
Physics, Sprin\-ger, New York, 1994); 
C. Ord\'{o}\~{n}ez, L. Ray, and U. van Kolck, Phys.\,Rev.\,Lett.\ 
{\bf
72},  1982 (1994); C.\,M. Shakin, Wei--Dong Sun, and J. Szweda,
Phys.\,Rev.\, C {\bf  52}, 3353 (1995).
%
\bibitem{jae96}
L. J\"ade and H.\,V. von Geramb, LANL e--print archive
nucl--th/9604002, submitted to Phys.\,Rev.\ C.
%
\bibitem{Mach89}
R. Machleidt, Adv.\ in Nucl.\,Phys.\ {\bf 19}, 189 (1989).
%
\bibitem{inv}
H. Kohlhoff and H.\,V. von Geramb, in {\it Quantum
Inversion Theory and Applications}, Proceedings of the 109th
W.\,E. Heraeus Seminar, Bad Honnef 1993, edited by H.\,V. von Geramb
(Lecture Notes in Physics, Springer, New York, 1994);
M. Sander, C. Beck, B.\,C. Schr\"oder, H.--B. Pyo,
H. Becker, J. Burrows, H.\,V. von Geramb, Y. Wu, and S. 
Ishikawa, in 
{\it Conference Proceedings, Physics with GeV--Particle Beams} 
(World  
Scientific, Singapore 1995).
%
\bibitem{Burt81}
P.\,B. Burt, {\it Quantum Mechanics and Nonlinear Waves} 
(Harwood Academic, New York, 1981).
%
\bibitem{fth}
C. Itzykson and J.\,B. Zuber, {\it Quantum Field Theory} 
(Mc.\ Graw--Hill, New York 1980).
%
\bibitem{Jae95}
L. J\"ade and H.\,V. von Geramb, LANL e--print server, 
nucl--th/9510061 (1995).
%
\bibitem{ppg}
M. Jetter and H.\,V. von Geramb, Phys.\,Rev.\ C {\bf 49}, 1832 (1994).
%
\bibitem{trit}
B.\,F. Gibson, H. Kohlhoff, and H.\,V. von Geramb, Phys.\,Rev.\ C {\bf
51}, R465 (1995).
%
\bibitem{na}
H.\,F. Arellano, F.\,A. Brieva, M. Sander, and H.\,V. von Geramb, to
appear in Phys.\,Rev.\ C {\bf 54} (1996).
%
\bibitem{nijm}
V. Stoks and J.\,J. de Swart,  Phys.\,Rev.\ C {\bf  48}, 792 
(1993).
%
\bibitem{mach}
R. Machleidt and G.\,Q. Li, LANL e--print server, 
nucl--th/9301019 (1993).
%
\bibitem{arn94}
R.\,A. Arndt, I.\,I. Strakovsky and R.\,L. Workman, Phys.\,Rev.\ C {\bf
50}, 2731 (1994).
%
\end{thebibliography}
\end{document}